%% file: main.tex
\documentclass[sigconf]{acmart}
\AtBeginDocument{%
 \providecommand\BibTeX{{%
    \normalfont B\kern-0.5em{\scshape i\kern-0.25em b}\kern-0.8em\TeX}}}
    
\renewcommand\footnotetextcopyrightpermission[1]{} 
\usepackage{hyperref}
\usepackage{amsmath,amsfonts}
\usepackage{algorithmic}
\usepackage{graphicx}
\usepackage{textcomp}
\usepackage{xcolor}

\definecolor{babyblueeyes}{rgb}{0.63, 0.79, 0.95}
\newcommand{\Hilight}{\makebox[0pt][l]{\color{babyblueeyes}\rule[-4pt]{\linewidth}{10pt}}}

\usepackage{times}
\usepackage{fullpage}

\usepackage{booktabs}   
\usepackage{subcaption} 

\usepackage[ruled,vlined,linesnumbered]{algorithm2e}
\SetAlFnt{\small\sffamily}

\usepackage{listings}
\usepackage{textcomp}
\usepackage{enumitem}
\usepackage{cleveref}
\usepackage{url}
\usepackage{multirow}

\usepackage{tabularx}
\usepackage{rotating}
\usepackage{lscape}

\crefformat{section}{\S#2#1#3} 
\crefformat{subsection}{\S#2#1#3}
\crefformat{subsubsection}{\S#2#1#3}

\usepackage{tikz}
\newcommand*\circled[1]{\tikz[baseline=(char.base)]{
            \node[shape=circle,draw,inner sep=0.5pt] (char) {#1};}}

\usepackage{adjustbox}
\usepackage{array}

\newcolumntype{R}[2]{%
    >{\adjustbox{angle=#1,lap=\width-(#2)}\bgroup}%
    l%
    <{\egroup}%
}

\definecolor{blueryb}{rgb}{0.01, 0.28, 1.0}

\author{Bogdan Alexandru Stoica}
\authornote{This research was performed while the author was at EPFL.}
\affiliation{%
  \institution{University of Chicago}}

\author{Swarup K. Sahoo}
\affiliation{%
  \institution{DeepFence Inc.}
}

\author{James R. Larus}
\affiliation{%
  \institution{EPFL}
}

\author{Vikram S. Adve}
\affiliation{%
  \institution{University of Illinois at Urbana-Champaign}}

\renewcommand{\shortauthors}{Stoica, et al.}

\begin{abstract}
Dynamic program slicing can significantly reduce the code developers need to inspect, by narrowing it down to only a subset of relevant program statements.
However, despite an extensive body of research showing its usefulness, dynamic slicing is still short from production-level use due to the high cost of runtime instrumentation.

As an alternative, we propose statistical program slicing, a novel hybrid dynamic-static slicing technique that explores the trade-off between accuracy and runtime cost.
Our approach relies on modern hardware support for control flow monitoring and a novel, cooperative heap memory tracing mechanism combined with static program analysis for data flow tracking.
We evaluate statistical slicing for debugging on $\nrBugs$ failures from $\nrApps$ widely deployed applications and show it recovers $\woktAvgRRsliceT$ of the program statements on a dynamic slice with only $\wokAvgOverhead$ overhead.
\end{abstract}

\begin{document}

\title{Statistical Program Slicing: a Hybrid Slicing Technique for Analyzing Deployed Software}


\input{params}

\input{abstract}

\maketitle

\input{intro}


\input{example}

\input{design}

\input{implementation}

\input{eval}

\input{discussion}

\input{relwork}

\input{conclusions}

\bibliographystyle{plain}
\bibliography{main}

\end{document}

%% file: params.tex
\newcommand{\wok}            {\textsc{Wok}}
\newcommand{\giri}           {\textsc{Giri}}

\newcommand{\nrApps}         {6}
\newcommand{\nrBugs}         {21}
\newcommand{\nrTracesI}      {214,798}
\newcommand{\nrTraces}       {3,873,616}

\newcommand{\wokAvgOverhead}					{5\%}

\newcommand{\wokfAvgRRsliceF}                   {100\%}
\newcommand{\wokfAvgRRrootcF}                   {96\%}
\newcommand{\wokfAvgSizeRatioD}		            {1.2\times}
\newcommand{\wokfAvgSizeRatioS}                 {31\times}
\newcommand{\woktAvgRRsliceT}                   {94\%}
\newcommand{\woktAvgRRrootcT}                   {92\%}
\newcommand{\woktAvgSizeRatioD}		            {2\times}
\newcommand{\woktAvgSizeRatioS}                 {21\times}

\newcommand{\avgRRdeps}                         {82\%}


\newcommand{\gzipAname}                         	{\tt gzip-46312} 
\newcommand{\gzipAgiriS}                 			{771}
\newcommand{\gzipAgiriD}             				{130}
\newcommand{\gzipAgiriRC}        					{60}
\newcommand{\gzipAwokF}              				{130 (100\%)}
\newcommand{\gzipAwokFRC}         					{60 (100\%)}
\newcommand{\gzipAwokT}              				{130$\;\;\:$(98\%)}
\newcommand{\gzipAwokTRC}         					{58$\;\;\:$(95\%)}
\newcommand{\aGzipPrecTotalDeps}                	{39}
\newcommand{\aGzipWokTotalDeps}                 	{41}
\newcommand{\aGzipTotalRatio}                   	{39 (100\%)}
\newcommand{\aGzipMissedRatio}                  	{0}
\newcommand{\aGzipFalseRatio}                   	{2}

\newcommand{\curlAname}                         	{\tt curl-965} 
\newcommand{\curlAgiriS}                 			{771}
\newcommand{\curlAgiriD}             				{130}
\newcommand{\curlAgiriRC}        					{60}
\newcommand{\curlAwokF}              				{130 (100\%)}
\newcommand{\curlAwokFRC}         					{60 (100\%)}
\newcommand{\curlAwokT}              				{130$\;\;\:$(98\%)}
\newcommand{\curlAwokTRC}         					{58$\;\;\:$(95\%)}
\newcommand{\aCurlPrecTotalDeps}                	{512}
\newcommand{\aCurlWokTotalDeps}                 	{493}
\newcommand{\aCurlTotalRatio}                   	{445$\;\;\:$(88\%)}
\newcommand{\aCurlMissedRatio}                  	{67}
\newcommand{\aCurlFalseRatio}                   	{48}

\newcommand{\cppcAname}                             {\tt cppchk-5780} 
\newcommand{\cppcAgiriS}                 			{43,904}
\newcommand{\cppcAgiriD}             				{628}
\newcommand{\cppcAgiriRC}        					{48}
\newcommand{\cppcAwokF}              				{1,718$\;\;\:$(99\%)}
\newcommand{\cppcAwokFRC}         					{49 (100\%)}
\newcommand{\cppcAwokT}              				{3,031$\;\;\:$(92\%)}
\newcommand{\cppcAwokTRC}         					{61$\;\;\:$(94\%)}
\newcommand{\aCppcheckPrecTotalDeps}                {712}
\newcommand{\aCppcheckWokTotalDeps}                 {1,617}
\newcommand{\aCppcheckTotalRatio}                   {529$\;\;\:$(75\%)}
\newcommand{\aCppcheckMissedRatio}                  {}
\newcommand{\aCppcheckFalseRatio}                   {}

\newcommand{\cppcBname}                             {\tt cppchk-5909} 
\newcommand{\cppcBgiriS}                 			{59,926}
\newcommand{\cppcBgiriD}             				{241}
\newcommand{\cppcBgiriRC}        					{45}
\newcommand{\cppcBwokF}              				{267 (100\%)}
\newcommand{\cppcBwokFRC}         					{45 (100\%)}
\newcommand{\cppcBwokT}              				{1,097$\;\;\:$(90\%)}
\newcommand{\cppcBwokTRC}         					{47 (100\%)}
\newcommand{\bCppcheckPrecTotalDeps}                {55}
\newcommand{\bCppcheckWokTotalDeps}                 {252}
\newcommand{\bCppcheckTotalRatio}                   {45$\;\;\:$(80\%)}
\newcommand{\bCppcheckMissedRatio}                  {}
\newcommand{\bCppcheckFalseRatio}                   {}

\newcommand{\cppcCname}                             {\tt cppchk-5950} 
\newcommand{\cppcCgiriS}                 			{65,123}
\newcommand{\cppcCgiriD}             				{1,914}
\newcommand{\cppcCgiriRC}        					{9}
\newcommand{\cppcCwokF}              				{1,951 (100\%)}
\newcommand{\cppcCwokFRC}         					{43 (100\%)}
\newcommand{\cppcCwokT}              				{3,245$\;\;\:$(92\%)}
\newcommand{\cppcCwokTRC}         					{48 (100\%)}
\newcommand{\cCppcheckPrecTotalDeps}                {612}
\newcommand{\cCppcheckWokTotalDeps}                 {1,195}
\newcommand{\cCppcheckTotalRatio}                   {507$\;\;\:$(83\%)}
\newcommand{\cCppcheckMissedRatio}                  {}
\newcommand{\cCppcheckFalseRatio}                   {}

\newcommand{\cppcDname}                             {\tt cppchk-6059} 
\newcommand{\cppcDgiriS}                 			{53,322}
\newcommand{\cppcDgiriD}             				{1,189}
\newcommand{\cppcDgiriRC}        					{10}
\newcommand{\cppcDwokF}              				{1,217 (100\%)}
\newcommand{\cppcDwokFRC}         					{36 (100\%)}
\newcommand{\cppcDwokT}              				{2,021$\;\;\:$(94\%)}
\newcommand{\cppcDwokTRC}         					{43$\;\;\:$(90\%)}
\newcommand{\dCppcheckPrecTotalDeps}                {219}
\newcommand{\dCppcheckWokTotalDeps}                 {724}
\newcommand{\dCppcheckTotalRatio}                   {173$\;\;\:$(79\%)}
\newcommand{\dCppcheckMissedRatio}                  {}
\newcommand{\dCppcheckFalseRatio}                   {}

\newcommand{\cppcEname}                             {\tt cppchk-6106} 
\newcommand{\cppcEgiriS}                 			{42,805}
\newcommand{\cppcEgiriD}             				{1,227}
\newcommand{\cppcEgiriRC}        					{26}
\newcommand{\cppcEwokF}              				{1,239 (100\%)}
\newcommand{\cppcEwokFRC}         					{26 (100\%)}
\newcommand{\cppcEwokT}              				{2,960$\;\;\:$(91\%)}
\newcommand{\cppcEwokTRC}         					{23$\;\;\:$(89\%)}
\newcommand{\eCppcheckPrecTotalDeps}                {158}
\newcommand{\eCppcheckWokTotalDeps}                 {218}
\newcommand{\eCppcheckTotalRatio}                   {173$\;\;\:$(78\%)}
\newcommand{\eCppcheckMissedRatio}                  {124}
\newcommand{\eCppcheckFalseRatio}                   {}

\newcommand{\sqltAname}                        		{\tt sqlite-be84e3}  
\newcommand{\sqltAgiriS}                 			{9,024}
\newcommand{\sqltAgiriD}             				{4,151}
\newcommand{\sqltAgiriRC}        					{320}
\newcommand{\sqltAwokF}              				{4,282$\;\;\:$(99\%)}
\newcommand{\sqltAwokFRC}         					{318$\;\;\:$(98\%)}
\newcommand{\sqltAwokT}              				{5,789$\;\;\:$(94\%)}
\newcommand{\sqltAwokTRC}         					{391$\;\;\:$(95\%)}
\newcommand{\aSqlitePrecTotalDeps}                	{1,024}
\newcommand{\aSqliteWokTotalDeps}                 	{1,240}
\newcommand{\aSqliteTotalRatio}                   	{841$\;\;\:$(82\%)}
\newcommand{\aSqliteMissedRatio}                  	{183}
\newcommand{\aSqliteFalseRatio}                   	{399}

\newcommand{\sqltBname}                           	{\tt sqlite-78c0c8} 
\newcommand{\sqltBgiriS}                 			{28,278}
\newcommand{\sqltBgiriD}             				{3,783}
\newcommand{\sqltBgiriRC}        					{178}
\newcommand{\sqltBwokF}              				{3,931$\;\;\:$(99\%)}
\newcommand{\sqltBwokFRC}         					{176$\;\;\:$(98\%)}
\newcommand{\sqltBwokT}              				{5,305$\;\;\:$(96\%)}
\newcommand{\sqltBwokTRC}         					{201$\;\;\:$(98\%)}
\newcommand{\bSqlitePrecTotalDeps}                	{852}
\newcommand{\bSqliteWokTotalDeps}                 	{1,061}
\newcommand{\bSqliteTotalRatio}                   	{735$\;\;\:$(87\%)}
\newcommand{\bSqliteMissedRatio}                  	{197}
\newcommand{\bSqliteFalseRatio}                   	{406}

\newcommand{\sqltCname}                        		{\tt sqlite-32b63d} 
\newcommand{\sqltCgiriS}                 			{28,045}
\newcommand{\sqltCgiriD}             				{2,035}
\newcommand{\sqltCgiriRC}        					{35}
\newcommand{\sqltCwokF}              				{2,160$\;\;\:$(99\%)}
\newcommand{\sqltCwokFRC}         					{40 (100\%)}
\newcommand{\sqltCwokT}              				{3,423$\;\;\:$(94\%)}
\newcommand{\sqltCwokTRC}         					{45$\;\;\:$(97\%)}
\newcommand{\cSqlitePrecTotalDeps}                	{781}
\newcommand{\cSqliteWokTotalDeps}                 	{1,094}
\newcommand{\cSqliteTotalRatio}                   	{648$\;\;\:$(83\%)}
\newcommand{\cSqliteMissedRatio}                  	{133}
\newcommand{\cSqliteFalseRatio}                   	{449}

\newcommand{\sqltDname}                        		{\tt sqlite-9f2eb3}        
\newcommand{\sqltDgiriS}                 			{26,534}
\newcommand{\sqltDgiriD}             				{3,972}
\newcommand{\sqltDgiriRC}        					{118}
\newcommand{\sqltDwokF}              				{4,089$\;\;\:$(99\%)}
\newcommand{\sqltDwokFRC}         					{120 (100\%)}
\newcommand{\sqltDwokT}              				{5,623$\;\;\:$(96\%)}
\newcommand{\sqltDwokTRC}         					{131$\;\;\:$(95\%)}
\newcommand{\dSqlitePrecTotalDeps}                	{916}
\newcommand{\dSqliteWokTotalDeps}                 	{1,208}
\newcommand{\dSqliteTotalRatio}                   	{770$\;\;\:$(86\%)}
\newcommand{\dSqliteMissedRatio}                  	{166}
\newcommand{\dSqliteFalseRatio}                   	{458}

\newcommand{\sqltEname}                           	{\tt sqlite-264b97} 
\newcommand{\sqltEgiriS}                 			{9,158}
\newcommand{\sqltEgiriD}             				{1,090}
\newcommand{\sqltEgiriRC}        					{26}
\newcommand{\sqltEwokF}              				{1,132$\;\;\:$(99\%)}
\newcommand{\sqltEwokFRC}         					{27 (100\%)}
\newcommand{\sqltEwokT}              				{1,913$\;\;\:$(91\%)}
\newcommand{\sqltEwokTRC}         					{25$\;\;\:$(88\%)}
\newcommand{\eSqlitePrecTotalDeps}                	{121}
\newcommand{\eSqliteWokTotalDeps}                 	{198}
\newcommand{\eSqliteTotalRatio}                   	{107$\;\;\:$(88\%)}
\newcommand{\eSqliteMissedRatio}                  	{ 5}
\newcommand{\eSqliteFalseRatio}                   	{91}

\newcommand{\clngAname}                        	    {\tt clang-25156} 
\newcommand{\clngAgiriS}                 			{536,170}
\newcommand{\clngAgiriD}             				{3,935}
\newcommand{\clngAgiriRC}        					{29}
\newcommand{\clngAwokF}              				{5,307$\;\;\:$(99\%)}
\newcommand{\clngAwokFRC}         					{25$\;\;\:$(78\%)}
\newcommand{\clngAwokT}              				{5,336$\;\;\:$(90\%)}
\newcommand{\clngAwokTRC}         					{23$\;\;\:$(74\%)}
\newcommand{\aClangPrecTotalDeps}                	{710}
\newcommand{\aClangWokTotalDeps}                 	{843}
\newcommand{\aClangTotalRatio}                   	{531$\;\;\:$(75\%)}
\newcommand{\aClangMissedRatio}                  	{179}
\newcommand{\aClangFalseRatio}                   	{312}

\newcommand{\clngBname}                             {\tt clang-28116} 
\newcommand{\clngBgiriS}                 			{565,205}
\newcommand{\clngBgiriD}             				{6,291}
\newcommand{\clngBgiriRC}        					{30}
\newcommand{\clngBwokF}              				{9,416$\;\;\:$(99\%)}
\newcommand{\clngBwokFRC}         					{27$\;\;\:$(90\%)}
\newcommand{\clngBwokT}              				{10,061$\;\;\:$(89\%)}
\newcommand{\clngBwokTRC}         					{34$\;\;\:$(83\%)}
\newcommand{\bClangPrecTotalDeps}                	{984}
\newcommand{\bClangWokTotalDeps}                 	{1,663}
\newcommand{\bClangTotalRatio}                   	{777$\;\;\:$(80\%)}
\newcommand{\bClangMissedRatio}                  	{207}
\newcommand{\bClangFalseRatio}                   	{886}

\newcommand{\clngCname}		                        {\tt clang-32638} 
\newcommand{\clngCgiriS}                 			{43,053}
\newcommand{\clngCgiriD}             				{5,545}
\newcommand{\clngCgiriRC}        					{53}
\newcommand{\clngCwokF}              				{6,716 (100\%)}
\newcommand{\clngCwokFRC}         					{50$\;\;\:$(87\%)}
\newcommand{\clngCwokT}              				{6,922$\;\;\:$(90\%)}
\newcommand{\clngCwokTRC}         					{47$\;\;\:$(81\%)}
\newcommand{\cClangPrecTotalDeps}	                {922}
\newcommand{\cClangWokTotalDeps}                 	{1113}
\newcommand{\cClangTotalRatio}                   	{718$\;\;\:$(78\%)}
\newcommand{\cClangMissedRatio}                  	{204}
\newcommand{\cClangFalseRatio}                   	{395}

\newcommand{\clngDname}                        		{\tt clang-33082}  
\newcommand{\clngDgiriS}                 			{498,586}
\newcommand{\clngDgiriD}             				{4,946}
\newcommand{\clngDgiriRC}        					{32}
\newcommand{\clngDwokF}              				{6,045$\;\;\:$(98\%)}
\newcommand{\clngDwokFRC}         					{26$\;\;\:$(90\%)}
\newcommand{\clngDwokT}              				{9,652$\;\;\:$(90\%)}
\newcommand{\clngDwokTRC}         					{25$\;\;\:$(87\%)}
\newcommand{\dClangPrecTotalDeps}                	{811}
\newcommand{\dClangWokTotalDeps}                 	{1019}
\newcommand{\dClangTotalRatio}                   	{631$\;\;\:$(78\%)}
\newcommand{\dClangMissedRatio}                  	{180}
\newcommand{\dClangFalseRatio}                   	{388}

\newcommand{\clngEname}                        		{\tt clang-33471}  
\newcommand{\clngEgiriS}                 			{568,555}
\newcommand{\clngEgiriD}             				{6,794}
\newcommand{\clngEgiriRC}        					{71}
\newcommand{\clngEwokF}              				{7,946$\;\;\:$(99\%)}
\newcommand{\clngEwokFRC}         					{65$\;\;\:$(90\%)}
\newcommand{\clngEwokT}              				{8,883$\;\;\:$(90\%)}
\newcommand{\clngEwokTRC}         					{66$\;\;\:$(91\%)}
\newcommand{\eClangPrecTotalDeps}                	{967}
\newcommand{\eClangWokTotalDeps}                 	{1303}
\newcommand{\eClangTotalRatio}                   	{777$\;\;\:$(80\%)}
\newcommand{\eClangMissedRatio}                  	{190}
\newcommand{\eClangFalseRatio}                   	{526}

\newcommand{\pythAname}                        		{\tt python-12608} 
\newcommand{\pythAgiriS}                 			{22,103}
\newcommand{\pythAgiriD}             				{9,073}
\newcommand{\pythAgiriRC}        					{229}
\newcommand{\pythAwokF}              				{9,167 (100\%)}
\newcommand{\pythAwokFRC}         					{216$\;\;\:$(92\%)}
\newcommand{\pythAwokT}              				{15,579$\;\;\:$(95\%)}
\newcommand{\pythAwokTRC}         					{232$\;\;\:$(92\%)}
\newcommand{\aPythonPrecTotalDeps}                	{2,376}
\newcommand{\aPythonWokTotalDeps}                 	{2,740}
\newcommand{\aPythonTotalRatio}                   	{2,051$\;\;\:$(86\%)}
\newcommand{\aPythonMissedRatio}                  	{325}
\newcommand{\aPythonFalseRatio}                   	{689}

\newcommand{\pythBname}                        		{\tt python-29028} 
\newcommand{\pythBgiriS}                 			{63,640}
\newcommand{\pythBgiriD}             				{7,226}
\newcommand{\pythBgiriRC}        					{724}
\newcommand{\pythBwokF}              				{7,431 (100\%)}
\newcommand{\pythBwokFRC}         					{727 (100\%)}
\newcommand{\pythBwokT}              				{14,274$\;\;\:$(96\%)}
\newcommand{\pythBwokTRC}         					{888$\;\;\:$(95\%)}
\newcommand{\bPythonPrecTotalDeps}                	{2,259}
\newcommand{\bPythonWokTotalDeps}                 	{2,704}
\newcommand{\bPythonTotalRatio}                   	{1,944$\;\;\:$(87\%)}
\newcommand{\bPythonMissedRatio}                  	{315}
\newcommand{\bPythonFalseRatio}                   	{760}

\newcommand{\pythCname}                        		{\tt python-27867} 
\newcommand{\pythCgiriS}                 			{63,704}
\newcommand{\pythCgiriD}             				{7,164}
\newcommand{\pythCgiriRC}        					{705}
\newcommand{\pythCwokF}              				{7,394 (100\%)}
\newcommand{\pythCwokFRC}         					{708 (100\%)}
\newcommand{\pythCwokT}              				{14,146$\;\;\:$(97\%)}
\newcommand{\pythCwokTRC}         					{865$\;\;\:$(99\%)}
\newcommand{\cPythonPrecTotalDeps}                	{2,233}
\newcommand{\cPythonWokTotalDeps}                 	{2,656}
\newcommand{\cPythonTotalRatio}                   	{1,930$\;\;\:$(86\%)}
\newcommand{\cPythonMissedRatio}                  	{303}
\newcommand{\cPythonFalseRatio}                   	{726}

\newcommand{\pythDname}                        		{\tt python-27945}  
\newcommand{\pythDgiriS}                 			{59,899}
\newcommand{\pythDgiriD}             				{7,076}
\newcommand{\pythDgiriRC}        					{39}
\newcommand{\pythDwokF}              				{7,289 (100\%)}
\newcommand{\pythDwokFRC}         					{43 (100\%)}
\newcommand{\pythDwokT}              				{18,094$\;\;\:$(95\%)}
\newcommand{\pythDwokTRC}         					{46$\;\;\:$(96\%)}
\newcommand{\dPythonPrecTotalDeps}                	{2,223}
\newcommand{\dPythonWokTotalDeps}                 	{2,650}
\newcommand{\dPythonTotalRatio}                   	{1,929$\;\;\:$(85\%)}
\newcommand{\dPythonMissedRatio}                  	{294}
\newcommand{\dPythonFalseRatio}                   	{721}

%% file: abstract.tex
\begin{abstract}
Dynamic program slicing can significantly reduce the code developers need to inspect by narrowing it down to only a subset of relevant program statements.
However, despite an extensive body of research showing its usefulness, dynamic slicing is still short from production-level use due to the high cost of runtime instrumentation.

As an alternative, we propose statistical program slicing, a novel hybrid dynamic-static slicing technique that explores the trade-off between accuracy and runtime cost.
Our approach relies on modern hardware support for control flow monitoring and a novel, cooperative heap memory tracing mechanism combined with static program analysis for data flow tracking.
We evaluate statistical slicing for debugging on $\nrBugs$ failures from $\nrApps$ widely deployed applications and show it recovers $\woktAvgRRsliceT$ of the program statements on a dynamic slice with only $\wokAvgOverhead$ overhead.
\end{abstract}

%% file: intro.tex
\section{Introduction}


Program slicing is a decomposition technique that extracts all program statements relevant to a particular \emph{seed} instruction $\sigma$.
Informally, program slicing answers the following question: ``\textit{Which instructions directly or indirectly affect the computations performed by $\sigma$?}''.
Initially proposed by Mark Weiser \cite{weiser81} as a mechanism for debugging, program slicing has been successfully applied to multiple areas of software engineering such as testing, compiler optimizations, security, software maintenance, or code integration \cite{tip94}.


Broadly, there are two main flavors of program slicing: \textit{static} and \textit{dynamic}.
\textit{Static slicing} makes conservative assumptions about which program statements affect the seed.
Specifically, this approach relies on static program analysis to determine \emph{any} program statement that can \emph{potentially} affect a target instruction regardless of the program execution.
This, in turn, causes static slicing to quickly lose precision, particularly when applied to programs that make extensive use of pointers \cite{zhang05}.
In contrast, \textit{dynamic slicing} identifies the minimal set of statements that affect the seed \emph{for a particular program run}~\cite{korel88,agrawal90}.
A dynamic slice is input-sensitive and uses program flow information captured at runtime.
This makes dynamic slicing a more precise alternative when analyzing real-world programs \cite{zhang03,zhang05}.

Dynamic slicing is an elegant and intuitive mechanism to reason about a particular program execution. 
A vast amount of research was been dedicated to improving the technique and implementing efficient slicing tools.
Significant improvements focused on precision and cost-effectiveness of dynamic slicing for in-house testing~\cite{zhang03,zhang06,xin09}.
Later, these algorithms were implemented as state-of-the-art dynamic slicing tools for source-level~\cite{swarup13,kasikci15} and machine-level code~\cite{reps16}. 
Other tools tackled slicing multi-language applications~\cite{binkley14} by iteratively removing statements while ensuring semantics preservation.
Several works proposed combining dynamic and static program analysis~\cite{mock05,wee10,kasikci17,devecsery18} or reconstruct slices at a coarser abstraction level~\cite{manu07,zhang06} in order to improve scalability.

\begin{figure*}[h!]
  \includegraphics[width=\linewidth]{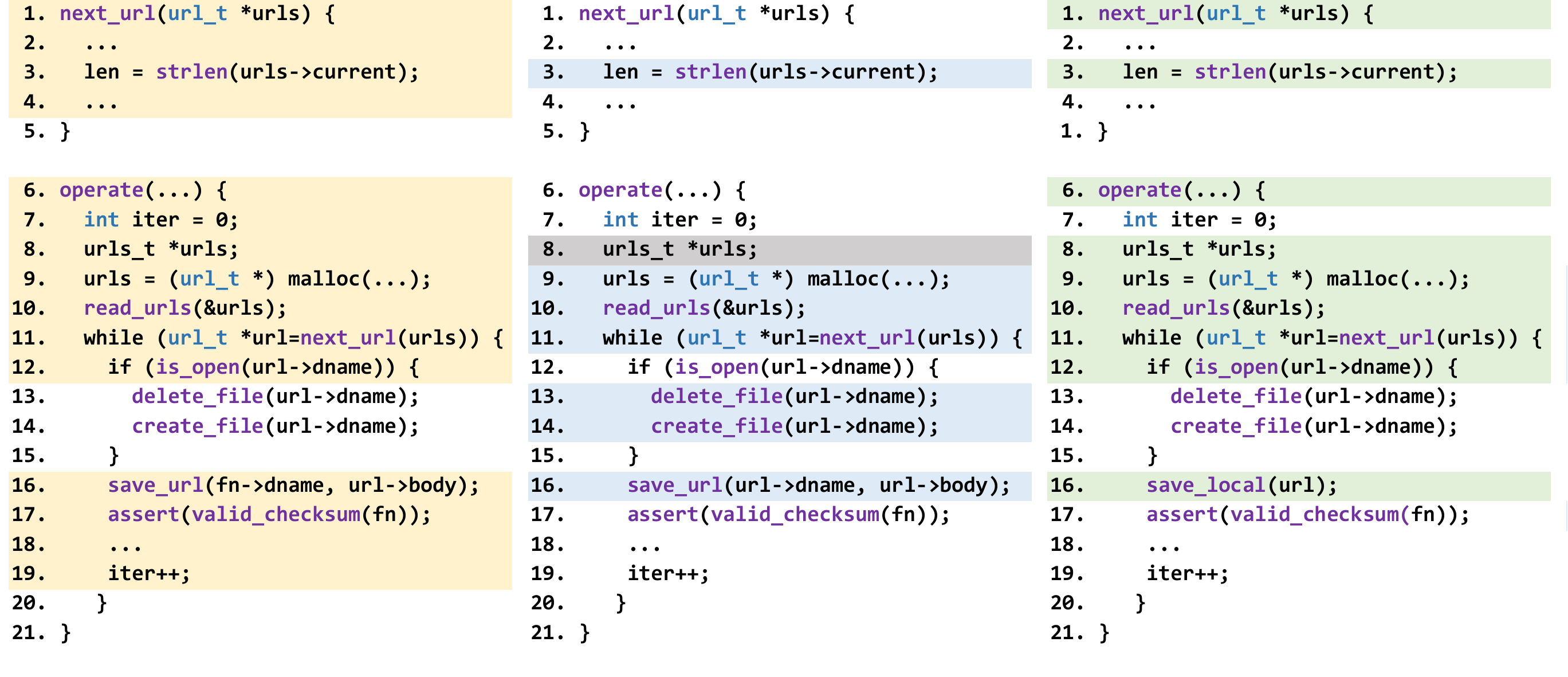}
  \vspace{-8mm}
  \caption{Example  of a statistical program slice (Curl bug \#965~\cite{curlbug}). The code snippet on the left highlights the data flow observed over multiple runs and depicts several data dependencies (blue arcs). The snippet in the center represents the control flow trace for the faulty run. The snippet on the right depicts a statistical slice after combing data and control flow observed at runtime, complemented by static program analysis.}
  \vspace{-5mm}
  \label{fig:fig0}
\end{figure*}

\subsection{Motivation}

Despite significant improvements over the past decades, dynamic program slicing is still unsuited for analyzing production failures~\cite{ernst19, devecsery18}.
In particular, developers face two main challenges in the context of today's modern software.

The first is related to the increasing difficulty of recovering faulty inputs in order to accurately reproduce production bugs offline.
Recent studies~\cite{zhang17,cui18,kasikci17} show that for an increasing number of user-site bugs, failure-inducing inputs are difficult to record either due to the high cost of capturing context and environment information~\cite{zhang17,cui18}, or because of privacy concerns~\cite{kasikci17}.

The second is associated with the high cost of the underlying control and data flow instrumentation for computing full dynamic slices which is unsuited for production use~\cite{devecsery18,kasikci17}. 
When failures cannot be reproduced offline because of missing or incomplete faulty inputs, developers default on runtime monitoring. This, in turn, creates a tension between deploying low-cost instrumentation and recording enough program behavior to reconstruct informative failure paths~\cite{orso15,parnin11,ernst19}.

We believe it is worthwhile to revisit dynamic slicing at its core and design algorithms that yield useful program slices at low (production-grade) cost, without burdening developers to obtain failure-inducing inputs.

\subsection{Contributions}

In this paper, we propose \textit{statistical program slicing}, a hybrid dynamic-static program slicing technique that explores the trade-offs between runtime overhead and slice accuracy.
The key insight of our technique is that we can reconstruct dynamic slices \emph{incrementally} from \emph{observed} program behavior through a combination of cooperative dynamic analysis and static program analysis.
Our approach is lightweight, with minimal effect on program execution; transparent, tracing executions without changes to the target binary; and, crucially, requires no knowledge of the faulty input. 

We implement statistical program slicing in a prototype called \wok\ (\cref{sec:design}).
\wok\ relies on Intel Processor Trace~\cite{intel17}---an efficient branch recording extension available on modern hardware---to collect precise control flow from faulty runs.
Since Intel PT lacks memory tracing capabilities, \wok\ further performs data flow analysis by using pointer tagging to strategically sample and track a subset of heap-allocated objects across multiple regular runs.
Finally, \wok\ relies on offline static alias analysis to complement dynamic data flow collection. 

Conceptually, \wok\ reconstructs an unsound approximation of the traditional program dependency graph~\cite{Horwitz1990} in an incremental fashion.
For brevity, we call this new program abstraction an \emph{observed dependency graph}.

\medskip

Although distributing data flow monitoring reduces the overhead to under $5\%$ (see \cref{sec:evaluation}), aggregating data dependencies over multiple runs poses an additional challenge:
We cannot reliably compare memory accesses between two distinct runs due to program relocation techniques (e.g., address space layout randomization).
To mitigate this issue, we discard memory address information and keep data dependencies as pairs of static program statements instead (similar to~\cite{agrawal90}).
We later partially recover control flow sensitivity by relying on the precise control flow from the faulty run collected by Intel PT and pruning data dependencies unrelated to the failure.
As a consequence, our slices represent a projection of the dynamic slice on static code. 
Thus, conceptually, statistical slices dwell in the space between static and dynamic slices. 

\medskip

We use \wok\ to reconstruct dynamic slices for $\nrBugs$ bugs from $\nrApps$ widely used open source applications.
We compare \wok's output against \giri, a state-of-the-art program slicing tool~\cite{swarup13}.
Specifically, we (1) measure how many program statements of the original dynamic slice can our tool recover, (2) the size of our resulting statistical slices, as well as (3) the accuracy of the recovered failure path (i.e., the set of instructions between the fault and the root cause).
Our evaluation (\cref{sec:evaluation}) shows that \wok\ recovers $\woktAvgRRsliceT$ of the unique program statements of the dynamic slice with only $5\%$ runtime overhead.
Overall, statistical slices are, on average, $89\%$ larger than original dynamic slices.
However, the corresponding failure paths are, on average, only $21\%$ larger.
We consider this is an acceptable trade-off between over-approximation and low overhead (i.e., $\leq 5\%$) that makes our technique suitable for production use.

To summarize, we make the following contributions:
\begin{itemize}[leftmargin=*]
\item[---] We present statistical program slicing, a hybrid dynamic--static slicing technique that leverages efficient hardware support for runtime monitoring combined with a customized static alias analysis to improve slice accuracy (\cref{sec:design});
\item[---] We develop a novel technique for runtime data flow analysis by relying on pointer tagging and distributing memory tracing across multiple runs of the same program (\cref{sec:design-dft});
\item[---] We evaluate our approach on $\nrBugs$ failures from $\nrApps$ widely deployed applications and show its usefulness for bug diagnosis. On average, we recover $\woktAvgRRsliceT$ of the program statements present on the corresponding dynamic slice with only $\wokAvgOverhead$ instrumentation overhead (\cref{sec:evaluation}).
\end{itemize}

%% file: example.tex

\section{Running Example}
\label{sec:example}

We illustrate statistical slices on an excerpt from Curl bug \#965~\cite{curlbug}. 
Curl allows users to fetch multiple URLs that share a common pattern (e.g., enumerating pages of a particular root domain) by using special symbols such as "\{\}"~\cite{curlurl}.
For instance, fetching pages from multiple ICSE editions can be encoded as "\texttt{curl -0 https://conf.researchr.org/home/icse-\{2019, 2020, 2021\}}". 
In this particular code version, having an unmatched bracket causes \texttt{urls->content} in function \texttt{next\_url} (line $3$) to become \texttt{NULL} which subsequently triggers a segmentation fault~\cite{curlbug}.

Figure \ref{fig:fig0} depicts the tree key stages of our algorithm: data flow tracking (left), control flow tracing (center) and combining the two together in a program slice (right),
While statistical slices are projected onto the source code, we collect control and data flow in the form of dependencies between program statements (blue and orange dotted arcs).
This helps us establish causality and decide which instructions to include in the final slice.

The snippet on the left highlights \wok's data flow tracking mechanism.
Our tool aggregates data flow from multiple runs to establish  dependencies between program statements accessing the same memory location.
For example, tagging \texttt{url} during allocation (line $9$), enables \wok\ can subsequently track all direct references of \texttt{url} (lines $11$, $14$, $18$) and derivatives (line $12$), and establish read/write dependencies between program statements (e.g., lines $11$ and $18$).

The snippet in the center illustrates control flow tracing via Intel PT.
While monitoring all runs, \wok\ retains control flow only when failures occurs.
In this particular example, the program statements highlighted in yellow show the code blocks executed when bug \#965 occurs.

The snippet on the right depicts the resulting statistical slice.
\wok\ combines data flow aggregated from multiple runs with the precise control flow information from the faulty run.
Data flow not exercised by the failure (e.g., line $14$) is discarded and control flow involving variables unrelated to the segmentation fault at line $3$ is disregarded (e.g., lines $8$, $9$, $20$, etc.).

Finally, we rely on static analysis to include program dependencies not captured by our runtime tracing strategies.
Typically these are relevant dependencies related to stack variables. 
In this particular example, static analysis helps us include the declaration line $7$ in the final slice.

%% file: design.tex


\section{Design}
\label{sec:design}

\wok\ operates in a client-server environment where runtime monitoring happens on the client (always-on) and slices are computed offline on the server whenever requested by developers.
Client-side, \wok\ relies on a combination of hardware support, sampling, and cooperative tracing to capture control and data flow at runtime.
Server-side, \wok\ combines execution information gathered cooperatively from multiple runs of the same program with customized static analysis, and computes a statistical slice on demand.
We thus expect \wok\ to run in a cooperative environment such as a datacenter or deployed on multiple user end-points, similar to prior failure path reconstruction tools \cite{ChilimbiLMNV09Holmes,kasikci15,liblit2007CBI,zhang17}.

\vspace{0.1cm}

\noindent \textbf{Client-side runtime monitoring (always-on)}. \wok\ monitors a target program on the client-side to record control and data flow information.
\wok\ tracks control flow using Intel Processor Trace, a new hardware extension which allows always-on execution monitoring at low overhead \cite{intel17} (\cref{sec:design-cft}).

As Intel PT has no memory monitoring capabilities, \wok\ tracks data flow by adding an extra level of indirection to memory addressing and leveraging the hardware protection mechanism available on x86-64 architectures (\cref{sec:design-dft}).
Specifically, \wok\ intercepts memory allocations and tags the high order bits thus making pointers uncanonical~\cite{intel17}.
This forces the CPU to raise a \emph{protection fault} and interrupt the target program every time a uncanonical address gets accessed.
When a trap occurs, \wok\ extracts information about the memory access (program counter, memory address, etc.) to infer data dependencies between program statements accessing the same memory location.
For brevity, we call this mechanism \emph{pointer poisoning}.

The key advantage of pointer poisoning is to enable monitoring for a range of addresses simultaneously and allows setting watchpoints at data-structure granularity.
Tagged bytes propagate virally every time the base pointer is copied or involved in pointer arithmetic.
One drawback is the overhead associated with hardware interrupts as every CPU fault triggers a context switch.
To reduce the cost, \wok\ samples a subset of the pointers to track per run over multiple executions.

Pointer poisoning is similar in spirit with behavioral watchpoints \cite{kumar13}.
However, there are several key differences.
Our mechanism generalizes to the entire heap and enables tracing heap memory arbitrarily.
Pointer poisoning further avoids the high cost of interrupts by selectively monitoring only a fraction of the pointers per run.
Moreover, the goal of our approach is to infer \verb|READ|--\verb|WRITE| dependencies between instructions that operate on the same bytes, as opposed to being switching dynamic binary translation on and off \cite{kumar13}.

\vspace{0.1cm}

\noindent \textbf{Server-side offline processing}. \wok\ computes statistical slices offline, on demand (e.g., after a failure occurs).
Our tool combines data flow collected cooperatively from multiple runs and precise control flow from the faulty execution into an \emph{observed dependency graph} where nodes represent static program statements and arcs represent control or data dependencies observed between them.
This graph is further augmented with statically inferred memory dependencies (e.g., stack variables) using a specialized static alias analysis restricted to the faulty code recorded by the PT trace (\cref{sec:design-aa}).

In order to compare and combine data flow from multiple runs, \wok\ keeps data dependencies as static pairs of program statements that accessed the same memory location thus discarding memory addresses, similar to~\cite{agrawal90}.
Thus, statistical slices represent a projected (i.e., flattened) version of their dynamic counterparts onto the source code (see Figure \ref{fig:fig0}).


\begin{figure}[!t]
  \centering \includegraphics[width=0.5\textwidth]{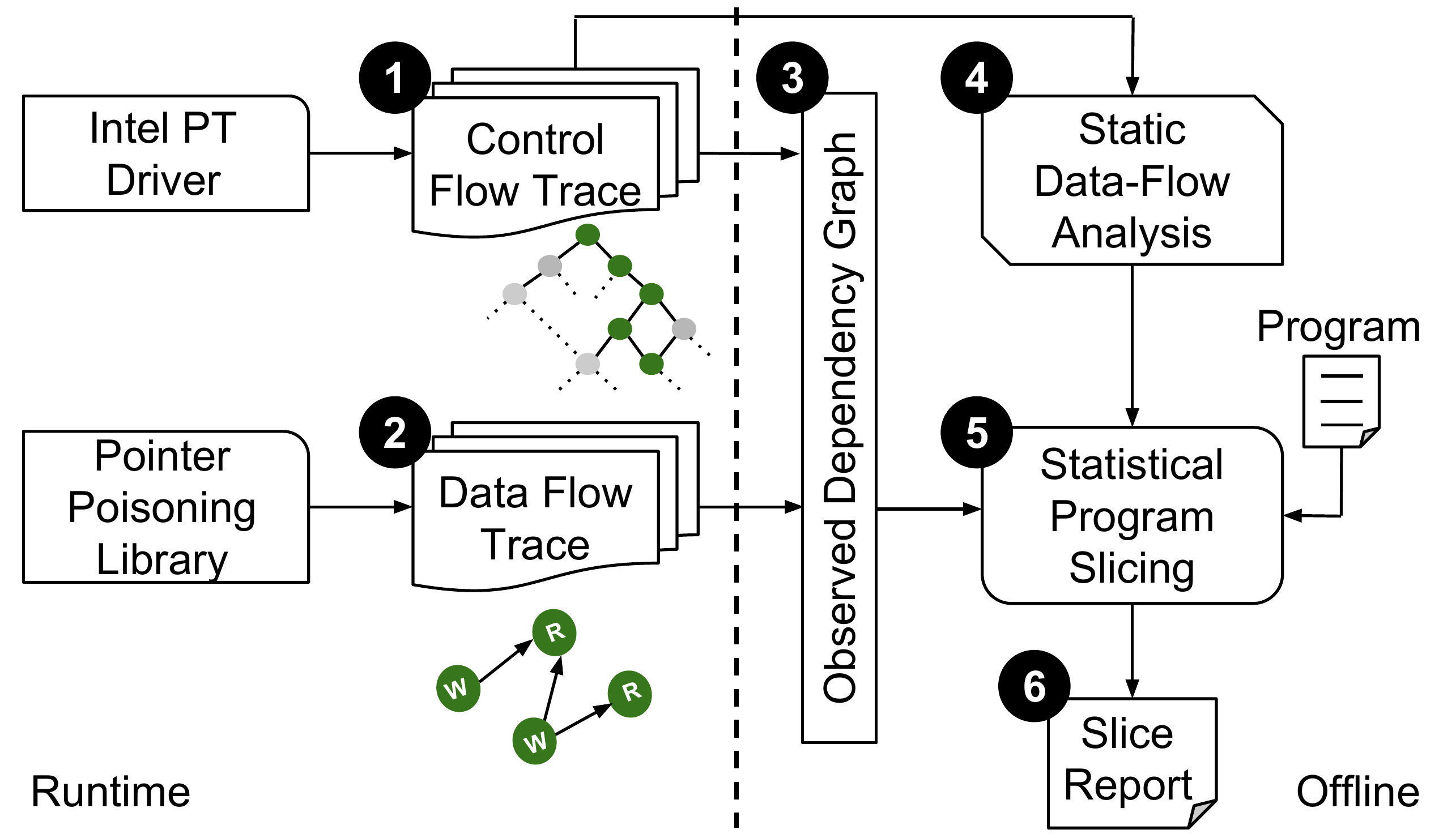} 
  \caption{\textbf{Overview of \wok's Architecture}: runtime monitoring phase (left) and slice crafting phase (right).}
  \label{fig:fig1-pps}
\end{figure}

\vspace{0.1cm}

\noindent \textbf{Usage model}. Figure \ref{fig:fig1-pps} illustrates the usage model of \wok.
Developers deploy two pieces of software on the client's machine: a shared library to perform pointer poisoning (\circled{1}, Fig. \ref{fig:fig1-pps}) and a kernel driver to enable the Intel PT tracing (\circled{2}).
When a failure occurs, developers instruct \wok\ to process the information collected at runtime and to compute a statistical program slice.
First, \wok\ combines the control flow and data flow into arcs in the observed dependency graph (\circled{3}).
Second, \wok\ augments the graph with statically inferred memory dependencies (e.g., stack variables) using a specialized static alias analysis restricted to faulty code only (\circled{4}).
Third, \wok\ performs a backward traversal on this augmented observed dependency graph starting from the seed instruction and walking backwards on observed dependencies (\circled{5}).
Lastly, developers are presented with a slice report detailing the static program statements that directly or indirectly affect the seed (\circled{6}).

\subsection{Data--Flow Monitoring}
\label{sec:design-dft}

The virtual address space on \texttt{x86-64} architectures can accommodate up to $2^{64}$ addresses (i.e., $16$ exabytes) which far exceeds the memory needs of modern applications
This far exceeds the needs of modern applications as current memory management implementations use only $48$ bits to represent memory addresses~\cite{intel17}.
Current systems us only $48$ bits to represent addresses.
The reminder high order $16$ bits are either all $0$ or $1$.
Departing from this specification makes the pointers uncanonical which raises a protection fault interrupting normal execution~\cite{intel17,kumar13}.

\begin{lstlisting}[caption=Pointer Poisoning Operations,label=lst:pp,escapechar=!][h!]
ptr = (unsigned int *)malloc(mySize);
...
// Pointer poisoning
!\Hilight!ptag = rand(0x0001, 0xfffe);
!\Hilight!ptr = ptr | (ptag << 48);
...
// Pointer unpoisoning
!\Hilight!ptr = ptr & 0x0000ffffffffffff;
*ptr = 42;
\end{lstlisting}

\noindent \textbf{Pointer Poisoning}. We take advantage of the memory protection mechanism described above to implement pointer poisoning.
We use a shared library to intercept heap allocation and tag base pointers.
Adding and removing tags are simply bitwise operations as illustrated in Listing \ref{lst:pp}.
Crucially, our approach makes no changes to the target binary and does not require recompilation.

\wok\ intercepts protection faults to peak at the current state of the execution.
Specifically, \wok\ records the program counter, memory location and type of memory access.
This information allows \wok\ to infer data dependencies between instructions that write and read the same address.
\wok\ preserves tags for the entire life time of the corresponding object in order to ensure continuous monitoring and preserve memory consistency.
Removing tags could further force us to maintain two versions of the same pointer--- tagged and un-tagged---and ensure compiler optimizations such as direct pointer comparisons work correctly.

The benefits of pointer poisoning are two-fold.
First, it allows us to trace memory virally: if $p$ is poisoned, its tag propagates whenever a new pointer is derived from $p$.
Second, unlike traditional hardware breakpoints that can trace a limited number of \emph{bytes} at a time, poisoning enables coarse grain memory monitoring at the \emph{object} level with potentially unlimited number of watchpoints. 

The drawback is the associated overhead penalty incurred by each protection fault.

\vspace{0.10cm}

\noindent \textbf{Selective Memory Instrumentation}. To reduce the overhead, we use poisoned pointers economically.
We strategically sample a subset of pointers to monitor per run.
This allows us to disseminate an expensive memory tracing procedure across multiple runs of the same program in a cooperative setting (e.g., a datacenter).



To avoid under-sampling infrequent code paths and increase coverage, we rely on an adaptive sampling strategy that ensures objects are sampled inversely proportional to their access frequency.
We rely on calling context (i.e., callstack) which is proven to be an effective predictor of objects with similar runtime behavior \cite{hauswirth04,novark09,sumner08} to differentiate between allocation contexts and re-adjust sampling rates.
Specifically, our policy oscillates between two phases:

\begin{itemize}[leftmargin=*]
\item[---] \emph{low frequency}: starting from $1.0$ (i.e., indiscriminate pointer poisoning), the sampling rate decreases exponentially if no memory accesses that lead to new data dependencies are being observed;
\item[---] \emph{high frequency}: once new memory accesses occur the sampling rate gets reset to $1.0$ and the process repeats.
\end{itemize}

We argue this best-effort strategy yields a reasonable approximation of the set of data dependencies executed during a faulty run and becomes more accurate with the number of runs observed.
Intuitively, memory objects are often accessed from a limited number of places in the source code~\cite{barr13} (e.g., a linked list is accessed through dedicated member functions).
Prior work has found that for short-running programs the set of memory accessing instructions is small and for long-running programs that set is mostly stable across different phases of the execution~\cite{novark09,chilimbi02}.
For this reason, we expect the number of data dependencies between instructions accessing memory objects to eventually converge.


\begin{figure}
  \centering \includegraphics[width=0.4\textwidth]{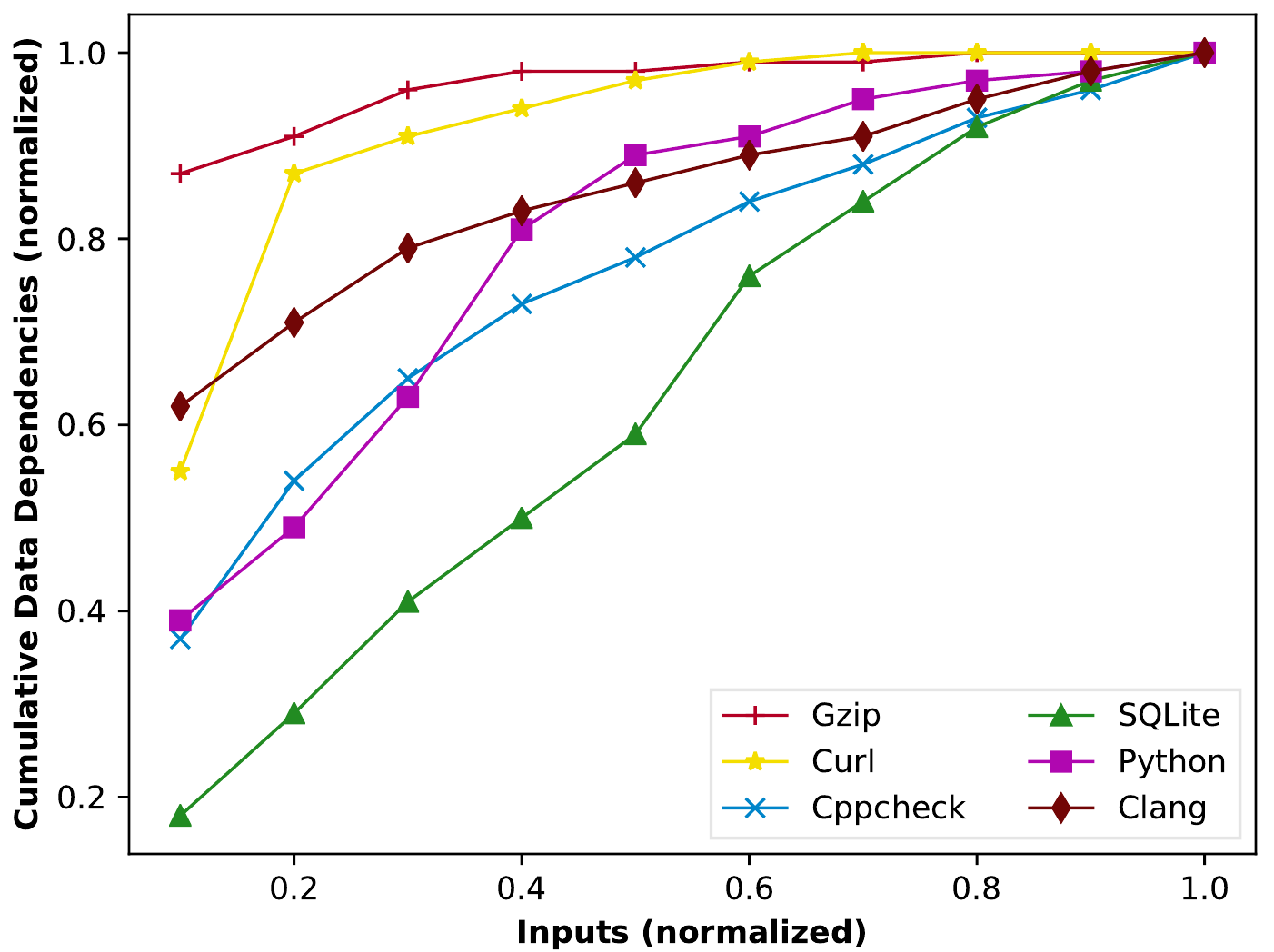} \caption{\textbf{Cumulative distribution of data dependencies}. Each cumulative distribution shows the fraction of the data dependencies recovered after each input. Each application runs against several thousands inputs.}
  \label{fig:fig2-cdf}
\vspace{-15pt}
\end{figure}

We observe similar behavior when analyzing memory access patterns for our evaluations (\cref{sec:evaluation}).
Figure \ref{fig:fig2-cdf} illustrates the cumulative distribution of data dependencies collected as a function of the number of runs.
As the number of runs increases, data dependency coverage begins to stabilize.
And a little over half the inputs account for $\sim$$80\%$ of the data dependencies for $5$ out of $\nrApps$.
SQLite's behavior is an artifact of the available test suite that include numerous inputs designed to maximize branch coverage and test functionality in isolation~\cite{th3}.

\subsection{Control Flow Monitoring}
\label{sec:design-cft}

We rely on Intel PT for runtime control flow tracing with low overhead.
Intel PT is an extension of the Intel PMU architecture available on Broadwell and Skylake CPU families \cite{intel17}.
Intel PT records control flow information in a highly compressed format ($\sim$$0.5$ bits per retired assembly instruction \cite{intel17}) which encodes all branches executed by the program.
Compressed traces are available for offline decoding to precisely recover the execution.
Moreover, Intel PT incurs a modest overhead of $3$-$4\%$ \cite{intel17,zhang17,kasikci15,kasikci17,ge2017}.


In our usage model (Fig. \ref{fig:fig1-pps}), \wok\ decodes Intel PT traces on the server-side.
Our tool uses this information to reconstruct precise control flow for the fault currently under investigation.
Precise control flow is a necessary ingredient for computing program slices.
Knowing which instructions executed during the faulty run helps \wok\ improve the accuracy of final slices in two ways.
First, it allows our tool to filter data flow collected in aggregate at runtime (i.e., in step \circled{2}, Fig. \ref{fig:fig1-pps}) by pruning data dependencies that were not exercised during the faulty execution. 
Second, it improves the precision of static alias analysis by restricting the scope of the procedure to executed code only (i.e., step \circled{4}, Fig. \ref{fig:fig1-pps}).




\subsection{Execution-driven Alias Analysis}
\label{sec:design-aa}

The main source of imprecision when computing statistical program slices comes from sampling. 
In addition, the data flow analysis described in \cref{sec:design-dft} focuses on heap memory (i.e., by tracking pointers) and, up until now, ignores data dependencies that can be inferred statically (e.g., those related to stack variables). 

We mitigate these shortcomings by using inter-procedural, flow-insensitive static alias analysis~\cite{andersen94}. 
Static alias analysis is known to be imprecise, especially in the presence of pointers \cite{weiser81,Korel1988,mock05,zhang05,kasikci13,nainar10,liblit03}.
To improve precision we restrict the scope of the procedure to code executed during the faulty run and retain only \emph{must} aliases.
Specifically, we use the PT trace to prune must aliasing pairs not exercised by the failure (\circled{4}, Figure \ref{fig:fig1-pps}) when iteratively computing points-to sets.
This allows us to perform an otherwise unscalable inter-procedural analysis using control flow information recorded by Intel PT for the faulty execution.

By disregarding may aliases, our procedure can potentially miss data dependencies for heap memory locations.
We take this conservative approach in order to preserve \emph{soundness} with respect to \emph{observed} program execution.
However, we will show that this restriction minimally impacts statistical slice accuracy in our evaluation (\cref{sec:evaluation}).


\subsection{Computing Statistical Program Slices}
\label{sec:design-staslicing}

\wok\ aggregates data dependencies collected cooperatively over multiple runs and combines them with the precise control flow information from the failing run to form an \emph{observed dependency graph}.
\wok\ transforms each program statement into a node and each control and data dependency observed at runtime into an arc.
To account for data dependencies potentially missed during sampling, \wok\ further complements this graph with must aliases arcs computed by the execution-driven static alias analysis described above (\ref{sec:design-aa}).
Finally, \wok\ performs a backward traversal on the observed dependency graph starting from the \emph{seed} instruction~\cite{horwitz98}.

We discard dynamic instance information for data dependencies since we cannot reliably order memory accesses from multiple runs.
Typically, ``flattening'' program flow can potentially lead to a significant size increase for the final slice size~\cite{agrawal90,zhang05}.
However, we mitigate most of these effects by pruning spurious data flow not exercised during the target execution (\cref{sec:evaluation}).


Algorithm \ref{alg:slicing} describes how \wok\ computes statistical program slices.
Initially, our prototype adds the $seed$ to a queue.
At each step it pops an unprocessed element from the queue and adds the relevant dependencies from the complemented observed dependency graph.
Using LLVM conventions \cite{lattner2004llvm} these dependencies can be either a conditional clause (e.g., \verb|if-then-else|), a function call/return, a function argument, or a memory access.


\begin{algorithm}[t!]
\caption{Statistical Slicing (simplified)}
\label{alg:slicing}

\DontPrintSemicolon
\SetAlgoLined
\LinesNumbered

$slice \gets \emptyset$ \;
$queue \gets \lbrace seed \rbrace$ \;
\While{$queue \neq \emptyset$}{
  $stmt \gets queue.pop()$ \;
  \uIf{$isConditional(stmt)$}{
      $cond \gets getCondition(stmt)$ \;
      $queue \gets queue \cup getDep(trace, cond)$ \;
  }
  \uElseIf{$isCall(stmt)$} {
      $ret \gets getRet(stmt)$ \;
      $queue \gets queue \cup getDep(trace, ret)$ \;
  }
  \uElseIf{$isArgument(stmt)$} {
      $args \gets getArgs(stmt)$ \;
      $queue \gets queue \cup getDep(trace, args)$ \;
  }
  \uElseIf{$readsMemory(stmt)$} {
      $queue \gets queue \cup getStores(stmt, obs\_dep\_graph)$\;
  }
  \uElseIf{$writesMemory(stmt)$} {
      $queue \gets queue \cup getWriteOpnd(stmt)$ \;
  }
  $queue \gets queue \cup getDep(stmt, trace)$ \;
  $slice \gets slice \cup stmt$ \;
}
\end{algorithm}

%% file: implementation.tex
%
\section{Implementation}
\label{sec:implementation}

We implement \wok\ using $2,643$ lines of C and C++ code for the runtime heap memory monitoring library, $342$ lines of Python code for the cooperative framework, $652$ lines of C code for the Intel PT driver \cite{kleen}, and $412$ lines of C++ code for the scope--restricted static alias analysis.
\wok\ is designed to be lightweight and transparent, making no changes to the underlying target application.
Rather, it uses a combination of function wrappers and the hardware memory protection mechanism (i.e., signals) in Linux for data flow tracing, and a loadable kernel driver to capture control flow.

\wok's static alias analysis and backward slicing (i.e., Algorithm \ref{alg:slicing}) are built on top of the LLVM framework \cite{lattner2004llvm}.
Consequently, \wok\ relies on LLVM's code analysis passes to infer static dependencies between instructions (e.g., def-use chains).
Since \wok\ is designed as a development tool, we expect it to have access to the source code of the target program.

\wok\ uses a dynamically loaded library to wrap heap memory functions (i.e., \texttt{malloc()} and \texttt{free()}) to poison and unpoison heap pointers.
\wok\ shifts valid pointers into the non--canonical address space by tagging their upper $4$ bytes to a non--zero value less than $65,535$ (\texttt{0xfffff}).
Thus, it employs the hardware memory protection mechanism which interrupts execution by issuing a \texttt{SIGSEGV} signal to be handled in user--space.
\wok\ intercepts the signal to patch the offending memory operand, record the instruction pointer (i,e., \texttt{RIP} register), the memory address and the type of access.
It later uses this information to establish \verb|READ|--\verb|WRITE| dependencies between instructions that operate on the same memory location.
To allow further monitoring, \wok\ preserves the tag by emulating in software the most frequently executed \texttt{x86-64} instructions (e.g., \texttt{mov}, \texttt{cmp}, \texttt{add}, etc.).
For the others, \wok\ lets the CPU execute the patched instruction and enables single--stepping (i.e., \texttt{TF} flag) to reapply the tag at the next instruction.
Note that we cannot reliably remove tags after an interrupt without (i) preventing our tool from monitoring future memory accesses made through the newly sanitized pointer and (ii) driving program execution into an inconsistent state by potentially keeping two versions of the same pointer in memory (e.g., tagged and non-tagged).

Intel PT tracing is enabled and disabled via a kernel module that uses the Machine State Register (MSR) interface.
Branches are stored in a $128MB$ hardware buffer that is occasionally flushed to RAM.
Intel PT can trace both the main binary and library code. 
For our evaluation purposes we focus on computing slices for the main binary only.

%% file: eval.tex

%

\section{Evaluation}
\label{sec:evaluation}



In this section we aim to answer the following research questions about statistical slicing: How accurate is our technique when compared to textbook program slicing and does it help developers reconstruct the failure paths for remote, hard-to-reproduce bugs (\ref{sec:eval-q1})? Also, can we compute statistical slices efficiently (\ref{sec:eval-q2})?

\subsection{Methodology}
\label{eval:method}

\noindent \textbf{Benchmarks}. To answer these questions, we evaluate \wok\ on $\nrApps$ widely utilized open-source applications: Clang \cite{clangurl}, a C/C++ compiler designed as drop-in replacement for GCC; SQLite \cite{sqliteurl}, an embedded database used in Chrome, Firefox, and Android; the Python interpreter \cite{pythonurl}; Cppcheck \cite{cppurl}, a C++ static analysis tool integrated with Visual Studio and Eclipse; Curl \cite{curlurl} a data transfer tool for FTP and HTTP protocols; and Gzip \cite{gzipurl}, a data compression tool. 

\vspace{0.10cm}

\noindent \textbf{Workloads}. We collect control and data flow information by running our test applications on a diverse set of inputs.
We select inputs that maximize the functionality exercised at runtime (i.e., coverage).
Our workloads include test suites, open-source and proprietary benchmarks~\cite{th3}, and regular application-specific inputs collected online (e.g., most $1,000$ popular C/C++ applications on Github).
In total, we track $\nrTraces$ executions at less than $5\%$ overhead each.

\vspace{0.10cm}

\noindent \textbf{Ground truth}. We evaluate \wok\ on $\nrBugs$ software failures randomly sampled from bug repositories or faults used in previous research studies~\cite{swarup13,kasikci15}.
To minimize bias, we consider bugs already fixed that can be reproduced.
We rely on the bug report to extract the seed instruction and the code patch that indicates the root cause as identified by the developer. 

We compare \wok against textbook static and dynamic program slices generated by \giri, a state-of-the-art program slicing tool~\cite{swarup13}.
We evaluate our technique in terms of recovery rate, slice size and the ability to faithfully reconstruct failure paths.



\vspace{0.10cm}

\noindent \textbf{Environment setup}. We run our experiments on a machine with to $8$-core Intel Xeon \texttt{E5-2680} CPU with Intel PT support and $32$ GB of RAM running Ubuntu 18.04.01 LTS with kernel version 4.15.0-58.


\subsection{Accuracy}
\label{sec:eval-q1}

We evaluate the accuracy of our technique on three different levels:

\input{tables/table-2}

\vspace{0.10cm}

\noindent \textbf{Recovery rate}. Given a precise dynamic slice as ground truth, we calculate what fraction of the program statements we actually recover.
Specifically, if $\sigma_{wok}$ and $\sigma_{giri}$ two sets of unique program statements produced by \wok\ and \giri\ respectively, we defined the recovery rate as the ratio between $\big| \sigma_{wok} \cap \sigma_{giri} \big|$ and $\big| \sigma_{giri} \big|$


\vspace{0.10cm}

\noindent \textbf{Slice size}. Statistical program slices live in the space between static and dynamic slices.
Thus, we further evaluate our slices in terms of size when compared to their textbook static and dynamic counterparts.


\vspace{0.10cm}

\noindent \textbf{Failure path reconstruction}. Finally, we evaluate how helpful statistical slices are during the fault diagnosis process.
Specifically, we measure the number of static program statements developers have to inspect in order to identify the root cause.
Similar to other hybrid code analysis tools~\cite{kasikci17,manu07,devecsery18}, we use a breadth-first traversal to explore instructions in increasing distance from the slicing seed (e.g., the smallest ``sphere''~\cite{binkley14,manu07})

Our evaluation works in three stages.
First, we run each test program against the training workloads described above.
Second, we compute statistical slices for each tested bug using the failing program statements as slice seeds.
Finally, we compare the output against the program slices produced by \giri.  

Typically, we expect \wok\ to be configured in an environment where it tracks a large and diverse set of inputs before computing slices.
For completeness, we also consider a different, albeit less frequent scenario where a failure with the same symptoms is encountered repeatedly~\cite{liblit05,kasikci15,ChilimbiLMNV09Holmes,kasikci13}.
Thus, instead of observing program behavior during normal utilization, we track data flow statistically from a particular faulty execution.

Note that our central findings, discussion and conclusions focus on the former, workload-driven scenario, rather than the latter.
However, allowing \wok\ to monitor the same faulty execution repeatedly is not without merit.  
Intuitively, our algorithm performs best when reconstructing slices using the minimal set of data dependencies necessary to compute precise dynamic slices.
Therefore, such a setup provides us with an ``upper bound'' on the accuracy and a ``lower bound'' on the size of statistical slices.
This also helps strengthen the hypothesis that as bug-related program flow coverage increases, \wok\ is able to better approximate dynamic slices (see Table \ref{table:t4-hddeps}).

Table \ref{table:t2-dynamic-slice} compares program slices, recovery rates and root cause diagnosis capabilities between \wok\ and \giri{}.
We break measurements down into slice sizes (columns labeled `\texttt{Slice}') and distance to the root cause (columns labeled `\texttt{Root cause}').
Recovery rates between statistical and dynamic slices are reported in brackets and calculated using the $\mu$ ratio described above.
For each failure, we instruct \wok\ to operate in both the workload-driven and failure only scenarios descried above.
Columns labeled `\wok\ --- \texttt{cooperative}' report numbers after collecting data flow information over the \emph{entire} set of training inputs.   
Columns labeled `\wok\ --- \texttt{failure only}' show measurements when operating exclusively over the set of data dependencies exercised during the failing run. 
Finally, we compare those against the program slices generated by \giri{} (columns `\giri{} --- \texttt{static}' and `\giri{} --- \texttt{dynamic}').

\wok\ recovers $\woktAvgRRsliceT$ of the program statements present on a precise dynamic slice (i.e., `\wok\ --- \texttt{cooperative}' columns).
\wok\ further recovers an average of $\woktAvgRRrootcT$ of the instructions linking the symptom to the root cause.
In terms of slice sizes, \wok's output is, on average, $89\%$ larger than that of \giri.
This is still $\woktAvgSizeRatioS$ smaller when compared to static slices.
While larger than textbook dynamic slices, we argue this is an acceptable trade-off between over-approximation and a $\wokAvgOverhead$ runtime overhead (\cref{sec:eval-q2}). 

\medskip

Our evaluation further reveals that having data flow exclusively from the faulty run enables \wok\ to recover almost $100\%$ of the corresponding dynamic slices. 
As expected, having $100\%$ failure-related data flow coverage translates into near-perfect dynamic slice recovery rate.
The few program statements missed are stack variable declarations which \wok\ is unable to detect through static program analysis.
We plan to mitigate such side effects by switching to a more powerful alias analysis.

\medskip

In part, missing program statements are caused by sampling.
The other cause is the quality of the static program analysis which is inherently imprecise~\cite{devecsery18,kasikci17}.
A close inspection revealed that the current alias analysis implementation excludes some stack variable allocation instructions.
However, we believe these to be the least critical components of a slice which developers can easily infer from the control flow portion of a statistical slice.

Ultimately, reconstruction accuracy is influenced by the ability of our tool to collect enough relevant data dependencies.
As coverage increases, our prototype is able to better approximate dynamic slices.


\medskip

Program slices are dependent upon the seed statements for which slicing is performed~\cite{zhang03}, potentially impacting measurements in Table \ref{table:t2-dynamic-slice}. 
To mitigate biases, we also devise a slice-independent evaluation and measure recovery rates between sets of data dependencies.
This is further motivated by the fact that the data flow instrumentation is the main source of approximation for statistical slices.

Table \ref{table:t4-hddeps} compares the set of dynamic data dependencies constructed by \wok\ and \giri{} respectively.
On average, \wok\ recovers $\avgRRdeps$ of the data dependencies located on the precise dynamic slices, while the recovery rate for full slices goes up to $\woktAvgRRsliceT$. 
This increase can be attributed to the number of program dependencies linking two statements (typically, $1$-$4$ in our measurements). 
Consequently, \wok\ needs only to capture only one of them to include the two instructions.

\input{tables/table-4}

\subsection{Efficiency}
\label{sec:eval-q2}

\begin{figure}[!t]
  \centering \includegraphics[width=0.5\textwidth]{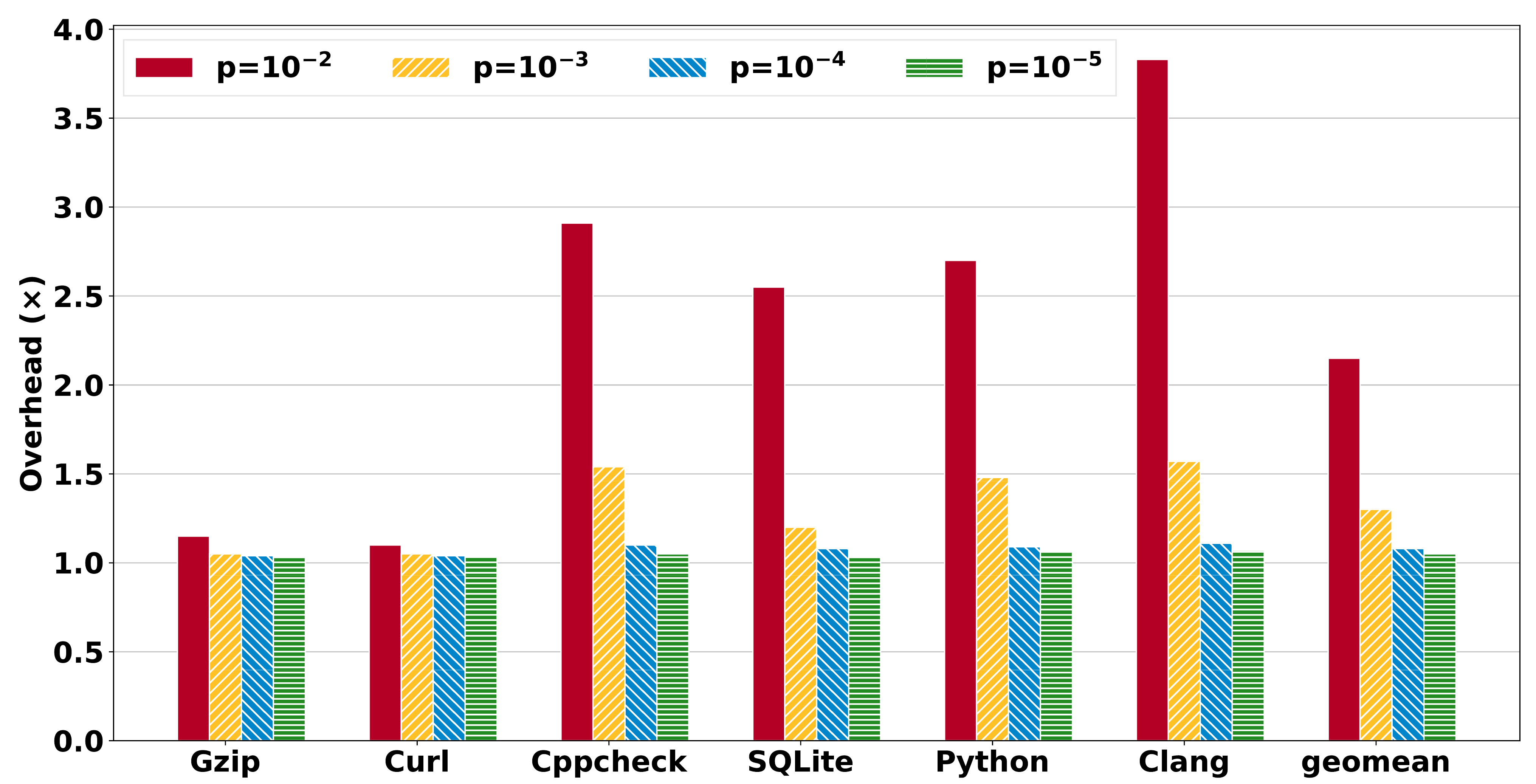}
  \caption{\textbf{Performance measurements}. Normalized overheads at different sampling rates ($p=10^{-1}$, $p=10^{-2}$...)}
  \label{fig:fig2-overhead}
\end{figure}

Figure \ref{fig:fig2-overhead} shows the runtime penalty incurred by \wok.
The overhead decreases with the sampling rate from $2.05\times$, $1.30\times$, $1.08\times$, to $1.05\times$ (geometric average).
Assuming an overhead budget of $\leq5\%$, we chose an optimal sampling rate between $\tfrac{1}{10,000}$ (i.e., $1$ in $10,000$ pointers allocated) and $\tfrac{1}{100,000}$.

At higher sampling rates, the ``lion's share'' of the overhead is due to the hardware interrupts triggered by dereferencing poisoned pointers.
On average, \wok\ incurs a penalty of $1.8$ $\mu$seconds per dereference.
At a close inspection we discovered that $50\%$ of the time is spent in kernel space, inside Linux's signal handling mechanism ($\texttt{do\_signal}$ -- $6.5\%$, $\texttt{sys\_rt\_sigreturn}$ -- $3.3\%$, and $\texttt{general\_protection}$ -- $2.7\%$).

At lower sampling rates, the overhead is dominated by Intel PT.
Intel PT incurs $5.8\%$ geometrical average slowdown with slight performance degradation for short running tasks (i.e., $\leq 1.0$ seconds).
This is due to the start-up and tear-down costs that cannot be amortized by the branch tracing itself.

Developers using our tool can perform a similar sensitivity analysis to determine which sampling rate to use.
The overhead of pointer poisoning for a given workload can be computed analytically as
\[ exp\_overhead = 1 + \frac{avg\_itrp\_cost \times exp\_hmem\_acc}{orig\_time} \]
where $orig\_time$ is the running time of the application without instrumentation, $avg\_trap\_cost$ is the average cost of handling a single interrupt, and $exp\_hmem\_acc$ is the expected number of heap memory accesses for the current workload.


%% file: tables/table-2.tex
\begin{table*}
\centering

\begin{tabular}{lrrrrrrr}
\multirow{2}{*}{\textbf{Bug ID}}   & \multicolumn{1}{r}{\multirow{2}{*}{\textbf{Giri --- static}}} & \multicolumn{2}{c}{\textbf{Giri --- dynamic}}                                    & \multicolumn{2}{c}{\textbf{Wok --- failure only}}                                          & \multicolumn{2}{c}{\textbf{Wok --- cooperative}}                                            \\
                                   & \multicolumn{1}{r}{}                                          & \multicolumn{1}{r}{\textbf{Slice}}   & \multicolumn{1}{r}{\textbf{Root cause}}   & \multicolumn{1}{r}{\textbf{Slice (\%)}}   & \multicolumn{1}{r}{\textbf{Root cause (\%)}}   & \multicolumn{1}{r}{\textbf{Slice (\%)}}   & \multicolumn{1}{r}{\textbf{Root cause (\%)}} \\ \hline \hline
\gzipAname                         &  \gzipAgiriS                                                  &  \gzipAgiriD                         &  \gzipAgiriRC                             &  \gzipAwokF                               & \gzipAwokFRC                                   & \gzipAwokT                                & \gzipAwokTRC                                 \\ \hline
\sqltAname                         &  \sqltAgiriS                                                  &  \sqltAgiriD                         &  \sqltAgiriRC                             &  \sqltAwokF                               & \sqltAwokFRC                                   & \sqltAwokT                                & \sqltAwokTRC                                 \\
\sqltBname                         &  \sqltBgiriS                                                  &  \sqltBgiriD                         &  \sqltBgiriRC                             &  \sqltBwokF                               & \sqltBwokFRC                                   & \sqltBwokT                                & \sqltBwokTRC                                 \\
\sqltCname                         &  \sqltCgiriS                                                  &  \sqltCgiriD                         &  \sqltCgiriRC                             &  \sqltCwokF                               & \sqltCwokFRC                                   & \sqltCwokT                                & \sqltCwokTRC                                 \\
\sqltDname                         &  \sqltDgiriS                                                  &  \sqltDgiriD                         &  \sqltDgiriRC                             &  \sqltDwokF                               & \sqltDwokFRC                                   & \sqltDwokT                                & \sqltDwokTRC                                 \\
\sqltEname                         &  \sqltEgiriS                                                  &  \sqltEgiriD                         &  \sqltEgiriRC                             &  \sqltEwokF                               & \sqltEwokFRC                                   & \sqltEwokT                                & \sqltEwokTRC                                 \\ \hline
\curlAname                         &  \curlAgiriS                                                  &  \curlAgiriD                         &  \curlAgiriRC                             &  \curlAwokF                               & \curlAwokFRC                                   & \curlAwokT                                & \curlAwokTRC                                 \\ \hline
\pythAname                         &  \pythAgiriS                                                  &  \pythAgiriD                         &  \pythAgiriRC                             &  \pythAwokF                               & \pythAwokFRC                                   & \pythAwokT                                & \pythAwokTRC                                 \\
\pythBname                         &  \pythBgiriS                                                  &  \pythBgiriD                         &  \pythBgiriRC                             &  \pythBwokF                               & \pythBwokFRC                                   & \pythBwokT                                & \pythBwokTRC                                 \\
\pythCname                         &  \pythCgiriS                                                  &  \pythCgiriD                         &  \pythCgiriRC                             &  \pythCwokF                               & \pythCwokFRC                                   & \pythCwokT                                & \pythCwokTRC                                 \\
\pythDname                         &  \pythDgiriS                                                  &  \pythDgiriD                         &  \pythDgiriRC                             &  \pythDwokF                               & \pythDwokFRC                                   & \pythDwokT                                & \pythDwokTRC                                 \\ \hline
\cppcAname                         &  \cppcAgiriS                                                  &  \cppcAgiriD                         &  \cppcAgiriRC                             &  \cppcAwokF                               & \cppcAwokFRC                                   & \cppcAwokT                                & \cppcAwokTRC                                 \\
\cppcBname                         &  \cppcBgiriS                                                  &  \cppcBgiriD                         &  \cppcBgiriRC                             &  \cppcBwokF                               & \cppcBwokFRC                                   & \cppcBwokT                                & \cppcBwokTRC                                 \\
\cppcCname                         &  \cppcCgiriS                                                  &  \cppcCgiriD                         &  \cppcCgiriRC                             &  \cppcCwokF                               & \cppcCwokFRC                                   & \cppcCwokT                                & \cppcCwokTRC                                 \\
\cppcDname                         &  \cppcDgiriS                                                  &  \cppcDgiriD                         &  \cppcDgiriRC                             &  \cppcDwokF                               & \cppcDwokFRC                                   & \cppcDwokT                                & \cppcDwokTRC                                 \\
\cppcEname                         &  \cppcEgiriS                                                  &  \cppcEgiriD                         &  \cppcEgiriRC                             &  \cppcEwokF                               & \cppcEwokFRC                                   & \cppcEwokT                                & \cppcEwokTRC                                 \\ \hline
\clngAname                         &  \clngAgiriS                                                  &  \clngAgiriD                         &  \clngAgiriRC                             &  \clngAwokF                               & \clngAwokFRC                                   & \clngAwokT                                & \clngAwokTRC                                 \\
\clngBname                         &  \clngBgiriS                                                  &  \clngBgiriD                         &  \clngBgiriRC                             &  \clngBwokF                               & \clngBwokFRC                                   & \clngBwokT                                & \clngBwokTRC                                 \\
\clngCname                         &  \clngCgiriS                                                  &  \clngCgiriD                         &  \clngCgiriRC                             &  \clngCwokF                               & \clngCwokFRC                                   & \clngCwokT                                & \clngCwokTRC                                 \\
\clngDname                         &  \clngDgiriS                                                  &  \clngDgiriD                         &  \clngDgiriRC                             &  \clngDwokF                               & \clngDwokFRC                                   & \clngDwokT                                & \clngDwokTRC                                 \\
\clngEname                         &  \clngEgiriS                                                  &  \clngEgiriD                         &  \clngEgiriRC                             &  \clngEwokF                               & \clngEwokFRC                                   & \clngEwokT                                & \clngEwokTRC                             
 \end{tabular}
\caption{\textbf{Statistical slicing evaluation}. Slice sizes (\texttt{Slice}) and distances to root causes measured in lines of code (\texttt{Root cause}) for static, dynamic and statistical slices. Recovery rates between statistical and dynamic slices are shown in brackets.}
\label{table:t2-dynamic-slice}
\end{table*}

%% file: tables/table-4.tex
\begin{table}[t!]
\centering
\begin{tabular}{lrrr}
\textbf{Bug ID} & \textbf{Giri} & \textbf{Wok} & \textbf{Overlap} (\%) \\ \hline \hline

\gzipAname & \aGzipPrecTotalDeps & \aGzipWokTotalDeps & \
\aGzipTotalRatio \\ \hline

\sqltAname & \aSqlitePrecTotalDeps & \aSqliteWokTotalDeps & \
\aSqliteTotalRatio \\

\sqltBname & \bSqlitePrecTotalDeps & \bSqliteWokTotalDeps & \
\bSqliteTotalRatio \\

\sqltCname & \cSqlitePrecTotalDeps & \cSqliteWokTotalDeps & \
\cSqliteTotalRatio \\

\sqltDname & \dSqlitePrecTotalDeps & \dSqliteWokTotalDeps & \
\dSqliteTotalRatio \\

\sqltEname & \eSqlitePrecTotalDeps & \eSqliteWokTotalDeps & \
\eSqliteTotalRatio \\
 \hline

\curlAname & \aCurlPrecTotalDeps & \aCurlWokTotalDeps & \
\aCurlTotalRatio \\ \hline

\pythAname & \aPythonPrecTotalDeps & \aPythonWokTotalDeps & \
\aPythonTotalRatio \\

\pythBname & \bPythonPrecTotalDeps & \bPythonWokTotalDeps & \
\bPythonTotalRatio \\

\pythCname & \cPythonPrecTotalDeps & \cPythonWokTotalDeps & \
\cPythonTotalRatio \\

\pythDname & \dPythonPrecTotalDeps & \dPythonWokTotalDeps & \
\dPythonTotalRatio \\ \hline

\cppcAname & \aCppcheckPrecTotalDeps & \aCppcheckWokTotalDeps & \
\aCppcheckTotalRatio \\

\cppcBname & \bCppcheckPrecTotalDeps & \bCppcheckWokTotalDeps & \
\bCppcheckTotalRatio \\

\cppcCname & \cCppcheckPrecTotalDeps & \cCppcheckWokTotalDeps & \
\cCppcheckTotalRatio \\

\cppcDname & \dCppcheckPrecTotalDeps & \dCppcheckWokTotalDeps & \
\dCppcheckTotalRatio \\

\cppcEname & \eCppcheckPrecTotalDeps & \eCppcheckWokTotalDeps & \
\eCppcheckTotalRatio \\ \hline

\clngAname & \aClangPrecTotalDeps & \aClangWokTotalDeps & \
\aClangTotalRatio \\

\clngBname & \bClangPrecTotalDeps & \bClangWokTotalDeps & \
\bClangTotalRatio \\

\clngCname & \cClangPrecTotalDeps & \cClangWokTotalDeps & \
\cClangTotalRatio \\

\clngDname & \dClangPrecTotalDeps & \dClangWokTotalDeps & \
\dClangTotalRatio \\

\clngEname & \eClangPrecTotalDeps & \eClangWokTotalDeps & \
\eClangTotalRatio

\end{tabular}
\vspace{0.05in} \hrule \vspace{0.025in}
\caption{\textbf{Slice--independent evaluation}. The size and overlap of the dynamic data dependency sets present on \giri{} and \wok's dependence graphs, repsectively.}
\label{table:t4-hddeps}
\end{table}

%% file: discussion.tex

%

\section{Discussion}
\label{sec:discussion}

In this section we address some remaining open questions.

\vspace{0.10cm}

\noindent \textbf{Usage scenarios:} We designed \wok\ to operate primarily in production distributed environment (e.g., a datacenter) where target applications can be monitored continuously across multiple executions.
The main goal of our work is to help developers analyze remote failures when they lack the fault-inducing inputs or alternative means to reproduce bugs locally.
While our approach addresses program slice without making assumptions about the type of failures or the utilization model, developers can use our tool in a variety of scenarios.

\wok\ can also be leveraged during in-house testing.
Our configurable sampling policy make it particularly suited for adaptively more aggressive data flow instrumentation.
Recent works~\cite{zhao17,cui18} suggests that programmers can tolerate up to $50\%$ performance degradation when using integrated test and build frameworks.

We envision statistical program slices as building blocks for additional program analysis for root cause diagnosis.
For instance, rank-based fault localization techniques achieve better precision when operating on a near-minimal set of program statements relevant to a failure~\cite{ernst19,swarup13,manu07}.

Alternatively, developers can also use our technique to perform more targeted monitoring by narrowing the scope of the underlying instrumentation with minimal manual intervention.
For example, instead of tracing memory indiscriminately, programmers can easily restrict pointer poisoning to a smaller subset of data structures (e.g., using type information).
Similarly, developers can chose which traces to include in the final observed dependency graph (e.g., faulty executions only).

\vspace{0.10cm}

\noindent \textbf{Definition of a root cause:} Providing a universal definition for root causes is still an open problem.
Bugs are fixed differently by different developers and can potentially exhibit multiple root causes.
Throughout this paper, we define the root cause as the set of program statements altered by the developer when patching the bug~\cite{swarup13}.
We believe this approach minimizes bias and allows us to compare \wok\ against a pre-established ground truth, independent from our technique.

\vspace{0.10cm}

\noindent \textbf{Fail-stop bugs:} \wok\ is a program slicing tool that requires an initial program statement to bootstrap its algorithm.
If the bugs investigated are not fail-stop, developers need to define custom failure models (e.g., assertions) to establish an appropriate slicing seed.
Like similar tools~\cite{liblit03,ChilimbiLMNV09Holmes,kasikci15,binkley14,madsen16,cui18}, our prototype cannot help analyze latent failures that silently corrupt program state without externally-observable effects.

\vspace{0.10cm}

\noindent \textbf{Availability of Intel PT:} Our prototype focuses on Intel architectures and relies on Intel PT, a hardware extension available on the Broadwell microarchitecture (i.e., from 2014 onwards~\cite{intel17}). 
However, statistical program slicing is not limited to a particular platform. 
Recent hardware-assisted control flow capabilities were added by other CPU manufacturers (e.g., ARM's EMT\cite{arm16}) and existing implementations enable their utilization within virtual environments~\cite{kwon18,schumilo17}.

\vspace{0.10cm}

\noindent \textbf{PT trace size:} For our evaluation, we configured Intel PT to record control flow in a $2$MB ring buffer which was sufficient to reconstruct slices with the accuracy and efficiency numbers reported.
This is far below the maximum buffer size of $128$MB supported by the current PT generation.
However, for long-running executions such a size may prove insufficient.
The alternative proposed by the original system designers is to periodically save the contents of the ring buffer to disk, with a modest performance penalty.
Recent works suggest that such cost gets amortized for long-running executions~\cite{cui18,ge2017}.

%% file: relwork.tex
%
%
\section{Related Work}
\label{sec:related}


\noindent \textbf{Program slicing}. Static slicing was initially introduced by Weiser~\cite{weiser81} as a program decomposition technique to help with bug diagnosis.
Ottenstein et al.~\cite{Ottenstein1984} refined the initial algorithm by recasting it as graph reachability problem. Horwitz et al.~\cite{Horwitz1990} further extended the approach by computing slices inter-procedurally.
Despite substantial improvements, static program slicing lacks adequate precision as it still relies on conservative program analysis to maintain soundness \cite{tip1994survey,zhang05}.

Korel et al.~\cite{korel88} were the first to suggest using dynamic program flow information and coined the term \emph{dynamic program slicing}. 
Agrawal et al.~\cite{agrawal90} explored various several trade-offs between precision and runtime overhead for this new technique.
Zhang et al.~\cite{zhang03,zhang05} developed more efficient dynamic slicing algorithms without sacrificing precision.
Although state-of-the-art, these techniques are prohibitively expensive and cannot be readily deployed in production.

Recent works use various strategies to prune spurious program statements~\cite{mock05,manu07,swarup13} or combine static and dynamic slicing for increased precision~\cite{wee10,devecsery18}.  
Mock et al.~\cite{mock05} improve accuracy by trimming static slices using dynamic points-to information.
Thin slicing~\cite{manu07} attempts to reduce the size of static slices by including \emph{producer statements} only, namely program dependencies relevant to track value flow.
Dual slicing~\cite{wee10} finds causal path for concurrent failures by alternating between slicing failing and successful executions.
Sahoo et al.~\cite{swarup13} implement \giri{} -- an efficient dynamic slicer based on \cite{zhang03} -- and use slicing to narrow down the set of potential root causes, offline.
Optimistic Hybrid Analysis~\cite{devecsery18} improves dynamic slicing by using an underlying unsound static analysis, yet disregards control--flow information.
These techniques are orthogonal to statistical program slicing and can be combined for additional benefits.

Statistical slicing can be viewed as an instantiation of observation-based slicing (ORBS) \cite{binkley14}.
Unlike ORBS, our approach operates at the program binaries, constructing slices by collecting program dependencies over multiple \emph{unmodified} runs, rather than iteratively deleting a maximum set of statements, recompile and check to retain execution behavior.

\medskip

\noindent \textbf{Adaptive failure path reconstruction}. Our work draws inspiration by the Cooperative Bug Isolation framework (CBI) \cite{liblit03,liblit2007CBI} and refinements~\cite{liblit05,nainar07,liu05,jiang07,chilimbi09,nainar10}.
Typically, CBI-based techniques achieve low overhead by using sparse sampling which, in turn, requires a failure to manifest tens or even hundreds of thousands of times to narrow down the root cause.
In contrast, statistical program slicing can help developers diagnose bugs by observing a particular failure only once.

Kasikci et al.~\cite{kasikci15} attempt to reconstruct the tail end of a dynamic slice relying on hardware breakpoints and Intel Processor Trace, a successor of Intel's Last Branch Record store~\cite{intel17}.
While significant improvements over CBI-based approaches, these techniques rely on the hypothesis that the distance between root causes and symptoms is small.
However, assuming short failure propagation paths has proven inaccurate for complex bugs~\cite{cui18}.

In a follow-up work, Kasikci et al.~\cite{kasikci17} attempt to correct this shortcoming by combining static program analysis with PT-based control flow monitoring to test pseudo--invariants based related to thread interleavings.
However, this improved approach is still restrictive as it is tailored for a particular subclass of concurrency bugs that follow a specific timing hypothesis.

Statistical program slicing makes no assumptions about the type of the bug or the length of the failure path. 

\medskip

\noindent \textbf{Hardware Memory Tracing}. Pointer poisoning is inspired from \emph{behavioral watchpoints}~\cite{kumar13}.
Similar to behavioral watchpoints, we tag the upper bytes of a pointer to track its behavior.
Unlike behavioral watchpoints, we collect information about which instructions access heap memory in user--space to infer read--write relationships between them.
We also instrument a small fraction of (user--space) pointers to instrument over multiple runs, in contrast to tagging only certain pointers from infrequently accessed data structures in the kernel.

Devietti et.al.~\cite{devietti12} also uses hardware instrumentation to monitor memory accesses.
However, their approach requires custom hardware support unavailable on commodity hardware.
In contrast, support for pointer poisoning (tagging) and hardware-assisted branch tracing are already available off-the-shelf on several platforms (e.g., Intel, ARM, AMD).

%% file: conclusions.tex
%
\section{Conclusions}
\label{sec:conclusions}

In this paper we presented statistical program slicing, a novel hybrid static--dynamic program slicing technique which leverages hardware support for control--flow tracing (Intel PT) and a cooperative selective heap memory tracking mechanism (pointer poisoning) for low overhead.
We described \wok, a tool that generates statistical program slices.

We tested \wok\ on $\nrBugs$ failures observed in production from $\nrApps$ real--world applications.
We showed that \wok\ efficiently recovers $\woktAvgRRsliceT$ of the statements typically present on a dynamic program slice and $92\%$ of the statements linking the symptom to the root cause with only $5\%$ instrumentation overhead.